\documentstyle[prl,aps,multicol,epsf]{revtex}

\begin{document} 

\draft 

\title{Imaging of Low Compressibility Strips in the Quantum Hall
Liquid} 

\author{G. Finkelstein, P.I. Glicofridis, S.H.
Tessmer\footnote{Present address: Michigan State University, East
Lansing, MI 48824}, R.C. Ashoori} 

\address{Department of Physics and Center for Material Science and
Engeneering, Massachusetts Institute of Technology, Cambridge, MA
02139}

\author{M. R. Melloch} 

\address{Department of Electrical Engineering, Purdue University, West
Lafayette, IN 37907}

\maketitle

\begin{abstract} 

Using Subsurface Charge Accumulation scanning microscopy we image
strips of low compressibility corresponding to several integer Quantum
Hall filling factors. We study in detail the strips at Landau level
filling factors $\nu =$ 2 and 4. The observed strips appear
significantly wider than predicted by theory. We present a model
accounting for the discrepancy by considering a disorder-induced
nonzero density of states in the cyclotron gap. 

\end{abstract}


\begin{multicols}{2} 


Many of the key features observed in the Quantum Hall effect may be
explained in terms of the transport through the quasi one-dimensional
edge channels\cite{Haug}. Each channel is formed where the energy of
the corresponding Landau level at the edge of the sample equals the
Fermi energy. Theorists predict that the edge channels should be
separated by narrow strips with precisely integer Landau level filling
factor\cite{Chang,Chklovskii}. In the model, the strips arise due to
the zero compressibility, {\it i.e.} $\frac{dn}{d\mu} =0$, of the 2DEG
at the cyclotron gap in the electron density of states between two
Landau levels. The formation of incompressible strips between the edge
channels drastically modifies their current equilibration\cite{Kolya}.
The microscopic structure of the strips may be directly revealed in
scanning experiments\cite{STM,McCormik,Amir}.
 
The recently-developed technique of Subsurface Charge Accumulation
(SCA) imaging measures the local charge accumulation in a 2D system
underneath a sharp scanning probe. This charge is driven by an AC
excitation applied to the 2D electron gas (2DEG), embedded inside a
semiconductor heterostructure. Capacitive coupling between the 2DEG
and the probe permits detection of the charge accumulated in the 2DEG
in-phase and $90 ^\circ$ lagging from the excitation. This technique
has proven successful in imaging local compressibility features in the
2DEG in the integer Quantum Hall regime\cite{STM}. In this paper, we
use SCA Microscopy to study low compressibility strips formed in a
presence of a smooth electron density gradient at different Landau
level filling factors. The strips are formed within the bulk of the
2DEG around an induced density perturbation. The measured strip widths
are a few times larger than that predicted by theory\cite{Chklovskii}.
We describe how our results may be explained if we take into account a
nonzero density of states in the cyclotron gap, caused by a
short-range potential. 

We study a standard 2DEG formed at the GaAs/AlGaAs interface $90 nm$
below the sample surface. It has an electron density of $n \approx
3.5\times 10^{11}/cm^{2}$ and mobility $\mu \approx 4 \times 10^{5}
cm^2 / Vsec$. A uniform doping layer $60 nm$ wide starts $10 nm$ below
the surface and is separated from the 2DEG by a $20 nm$ undoped spacer
layer. To create a density gradient in the 2DEG, we locally perturb
the charge distribution in the sample by applying a voltage of +2 to
+3.5 V between the scanning probe and the sample for $\sim 30$ sec. We
find that injecting the tunneling current results in an increase of
the 2DEG density in a region extending laterally a few microns from
the tunneling site. The electron density at the center of the
perturbation is typically about $20 \%$ higher than the bulk value.
After the tunneling current is switched off, this density profile does
not change with time. We speculate that the local modification of the
electron density is a result of electron transfer from the donor layer
to the 2DEG, similar to the persistent photoconductivity
effect\cite{Lang}.

To determine quantitatively the extent of the perturbation, we perform
Kelvin probe\cite{Kelvin} imaging of the perturbed region. In this
measurement, we mechanically vibrate the sample in the vertical
direction with a frequency of $2 kHz$ and an amplitude of $\sim 10 nm$
and measure the oscillating charge induced on the scanning probe. This
charge results from the electrostatic potential difference $\Delta V$
between the probe and the sample. It is proportional to $\Delta V
\frac{dC}{dz}$, where $C$ is the probe-sample capacitance and $z$ is
their separation. We calibrate the sensitivity of this measurement to
the potential difference by changing the voltage between the 2DEG and
the probe and measuring the resulting change in the signal. We then
place the scanning probe well outside the perturbed region and balance
the Kelvin signal by applying a DC voltage between the 2DEG and the
scanning probe to null the electric field created by the work function
difference between them. The signal reappears upon placing the
scanning probe above the perturbed region. Using the known sensitivity
of our measurement to the potential difference, we map the
electrostatic potential by scanning just above the sample surface. 

\begin{figure}
\epsfxsize=\linewidth
\epsfbox{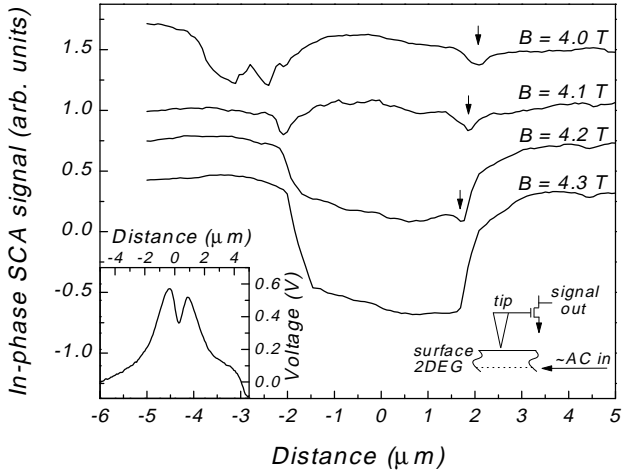}
\caption{\narrowtext
The SCA measurements at different magnetic fields
around $\nu =4$ across the area of enhanced density. The curves are
offset for clarity. Left inset: the Kelvin probe signal. Right inset:
a schematic description of the SCA technique. }
\end{figure}

The left inset of Fig. 1 shows the result of such a measurement. We
choose the origin of coordinates at the center of the perturbation and
fix the coordinate $y=0$. The signal is proportional to the deviation
in electron density from the unperturbed value and reaches its maximum
of $\Delta V \sim 0.5$ V close to the center of the perturbed region
(we ignore here a small dip in the Kelvin signal near $x=0$). Further
below we extract the 2DEG density gradient from the SCA measurements
in magnetic field to be $\frac {dn}{dx}= 5\times 10^{10}/cm^{2}\mu m$
at $x \approx 2\mu m$. We may also roughly estimate the density
enhancement in the 2DEG if we assume that the Kelvin signal results
from the vertical transfer of electrons from the donors to the 2DEG:
$\Delta V= 4 \pi e s \Delta n  /\epsilon$, where $s \sim 50 nm$ is the
distance between a typical donor in the doped layer and the 2DEG. From
the spatial derivative of the Kelvin voltage $\frac {dV}{dx} \sim 0.15
V/\mu m$ (left inset Fig. 1) we get $\frac {dn}{dx} = \frac {dV}{dx}
\epsilon /4\pi e s \sim 2\times 10^{11}/cm^{2} \mu m$, about four
times larger the result of the SCA measurements. Including the
possibility that the electrons may move to the 2DEG from the sample
surface reduces the estimate by a factor of two. Also, some of the
charge may be transferred vertically within the donor layer and not
from the donor layer to the 2DEG. These electrons contribute to the
Kelvin signal, but not to the 2DEG density, which further diminishes
the estimated $dn \over dx$. At present, we cannot determine the
precise charge balance within the various layers of the sample
contributing to the Kelvin probe signal.

We now describe SCA imaging\cite{STM}. The technique is depicted
schematically in the right inset of Fig. 1. In this measurement we
apply a small AC excitation (typically 3mV RMS in the linear response
regime) at a frequency of 10 -- 100 kHz to an ohmic contact at the
edge of the 2DEG. Due to the self-capacitance of the sample, a
resulting electric charge flows in and out from the 2DEG. We monitor
this charging of the 2DEG locally by placing a sharp scanning probe at
a small distance ($\sim 10 nm$) from the sample surface. The charge
induced on the scanning probe is proportional to the change in the
2DEG density. We measure the charge using a sensitive cryogenic charge
amplifier. The contrast in the SCA images indicates that different
regions of the 2DEG charge differently, due to variations in either
local compressibility or resistivity. Below, we discuss both
situations.

The SCA measurement performed at zero magnetic field does not show any
structure in the region of study, despite the induced perturbation.
Indeed, the 2DEG fully charges according to its self-capacitance and
the applied voltage. The situation changes when we apply a large
magnetic field. The SCA images at several different magnetic fields
are presented on a gray scale in Fig. 2. In magnetic fields around B=8
T ($\nu = 2$ in the bulk of the sample is reached at $B \approx 7$ T)
a ring feature appears in the SCA image. The magnitude of the SCA
signal is higher inside and outside the ring than on the ring itself.
As the field increases, the ring shrinks, moving toward the center of
the perturbed area. The results of a similar SCA measurement near $\nu
= 4$ on a different region of the sample (prepared in the same manner)
are shown in Fig. 1. Here we present only the signal measured on a
single line ($y=0$) across the diameter of the ring. As in Fig. 2, we
may trace the shrinkage of the ring as the field increases (see
feature marked by arrows, upper curves of Fig. 1). At even higher
magnetic fields the charging signal in the interior of the ring drops,
forming a circular depression in the SCA signal (Fig. 1, two lower
curves; Fig. 2c). 

If the scanning probe were located just above an incompressible
region, the 2DEG below would not charge and discharge with the weak AC
excitation. Thus in a region of low compressibility we detect a SCA
minimum\cite{comment}. We attribute the SCA minimum (Fig. 1) at B=4 T
to a low compressibility strip formed in the region of filling factor
$\nu = 4$. At higher field $\nu = 4$ corresponds to a higher electron
density causing the ring to move up the density gradient and shrink
toward the center. The position of the $\nu = 4$ strip shifts by
$\approx 200 nm$ as we step the magnetic field by 0.1 T. At $\nu = 4$,
this change in magnetic field corresponds to a $1\times
10^{10}/cm^{2}$ change in the electron density. Therefore, we estimate
the magnitude of the electron density gradient $\frac{dn}{dx}\approx
5\times 10^{10}/cm^{2} \mu m$ in the case of Figs. 1 and 3. The value
of the density gradient depends on the specific realization of the
density perturbation. In particular, in the region imaged at Fig. 2,
it is five times smaller, $\frac{dn}{dx}\approx 1\times 10^{10}/cm^{2}
\mu m$.

\begin{figure}
\epsfxsize=\linewidth
\epsfbox{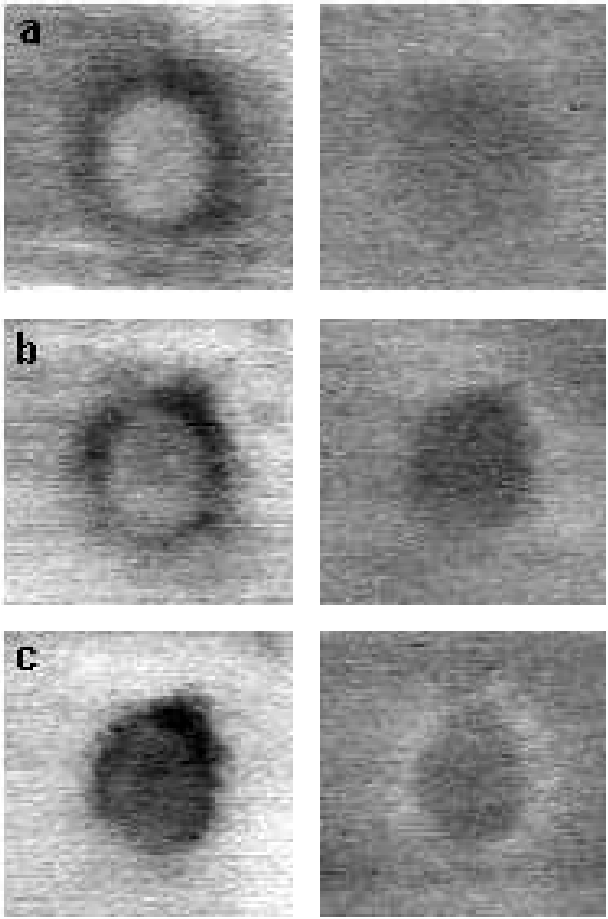}
\caption{\narrowtext
$13 \times 13 \mu m$ SCA images at different magnetic
fields near $\nu=2$ presented on a gray scale. The left and right
images correspond to the in-phase and $90 ^\circ$ lagging SCA signals,
respectively. The measurements are performed at magnetic fields of a)
$B=8.0$ T, b) $B=8.1$ T and c) $B=8.2$ T. Fig. 2a and c represent the
limiting cases where the SCA signal is dominated by compressibility or
resistivity, respectively. The data are taken at a different location
than Fig. 1.}
\end{figure}

A suppression of the high frequency SCA signal at a certain region
might be a result of either low compressibility or low conductivity.
In the latter case, there is insufficient time to charge the region
during the period of excitation. As the 2DEG longitudinal conductivity
is very low in the vicinity of integer filling factors, this
possibility must be examined carefully. To distinguish between the two
mechanisms we study both the in-phase and $90 ^\circ$ lagging SCA
signals as a function of frequency.

A negligible phase shifted signal and no frequency dependence were
observed at the conditions of Fig. 1, upper curve. This is evidenced
by the upper inset of Fig. 3, where the in-phase and $90 ^\circ$
lagging SCA signals are shown for the same conditions at frequencies
of 10, 30 and 100 $kHz$. We therefore conclude that the observed
feature reflects a suppressed 2DEG compressibility at integer filling
factor. However in some other situations we do observe a phase shifted
signal. Typically, at higher magnetic fields the SCA features appear
both in-phase and $90 ^\circ$ lagging from the excitation, this
demonstrating that the 2DEG does not have enough time to fully charge
during the excitation cycle. 

Under conditions giving rise to a phase shift, a ring-shaped feature
in the $90 ^\circ$ lagging signal accompanies the circular-shaped
region of the suppressed in-phase SCA signal (Fig. 2c). We interpret
this pattern as resulting from incomplete charging of the interior of
the circle through a poorly conducting integer $\nu$ strip around it.
This situation resembles an $RC$ circuit with a variable $R$ (lower
inset, Fig. 3). Moving the scanning probe toward the regions in the
interior of the strip increases the effective resistance. This causes
the measured in-phase signal to steadily decrease to zero, while the
$90 ^\circ$ lagging signal first increases from zero and then
decreases back to zero level. In such circumstances, it is difficult
to extract separately the conductivity and compressibility information
from the signal. We have performed extensive numerical modeling
allowing us to reproduce the major features observed in Fig.
1\cite{long}.

We use the lowest curve of Fig. 1 to estimate the sensitivity of the
SCA measurement. At this magnetic field the resistive strip at $\nu =
4$ prevents charging of the interior region, while the exterior region
charges fully\cite{comment}. This difference in signal levels between
the fully charging and non-charging regions may be taken as the
measure of the degree of charging. We observe that the contrast in the
low compressibility strip (Fig. 1 upper curve) reaches only about
$15\%$ of this value. Partially, the amplitude of the feature is
reduced due to nonlocal response of the scanning probe, which is
limited by the distance between the 2DEG and the surface. We estimate
the largest possible ``smearing'' from the sharpest features 
observed in the experiment. Taking into account the resulting response
function increases the actual depth of the SCA minimum at the $\nu
= 4$ strip only up to $25 \%$. We conclude that the 2DEG in the $\nu =
4$ strip remains partially compressible, and the large
geometric capacitance causes charge to enter this region.

The SCA signal is proportional to the capacitance between the tip and
the 2DEG. We crudely estimate the density of states (DOS) between
Landau levels $D$ by approximating the scanning probe and the 2DEG as
a parallel plate capacitor\cite{comment2}. The signal at the strip
location is $S(x)\propto 1/[d+\epsilon h+\epsilon (4\pi e^{2}
D)^{-1}]$, where $d = 90 nm$ is the 2DEG depth and $h = 10 nm$ is the
distance between the scanning probe and the surface\cite{Stern}. This
accounts for the 25 $\%$ decrease of the signal at the strip compared
to the neighboring regions, where $S(x)\propto 1/[d+\epsilon h]$. The
resulting $D$ is $\sim 30$ times smaller than the zero magnetic field
DOS. This result agrees with the values obtained from the
measurements of the DOS between Landau levels performed on bulk 2DEG
samples with lateral dimensions of $\sim 100 \mu m$\cite{DOS}. Our
observation demonstrates that the nonzero DOS in the cyclotron gap is
established on length scales smaller than the strip width.

The low compressibility strip discussed above for the filling factor
$\nu = 4$ reappears at filling factors $\nu = 2$ and $6$ (not shown).
We compare the strips at $\nu = 2$ and $4$ in Fig. 3. Their measured
widths are at least $w \sim 0.6 \mu m$ and $\sim 0.4 \mu m$,
respectively, as extracted by the eye from the flat portion of the SCA
minima. These values represent conservative estimates; detailed curve
fitting yields widths up to $30\%$ greater\cite{long}. The measured
widths are significantly larger than predicted by the
theory\cite{Chklovskii}. Indeed, taking $\frac{dn}{dx}\approx 5\times
10^{10}/cm^{2} \mu m$ from Fig. 1 we obtain from Eq. 20 of
Ref.\cite{Chklovskii} the widths $w_{0}$ of $0.23 \mu m$ and $0.17 \mu
m$ at $B = 8$ T ($\nu =2$) and $B = 4$ T ($\nu =4$), respectively. In
Fig. 2a the deviation is even more pronounced: the measured width of
$2 \mu m$ is four times larger than expected. We proceed now to show
how this discrepancy may be qualitatively explained by considering a
finite DOS localized between Landau levels\cite{comment3}. 

\begin{figure}
\epsfxsize=\linewidth
\epsfbox{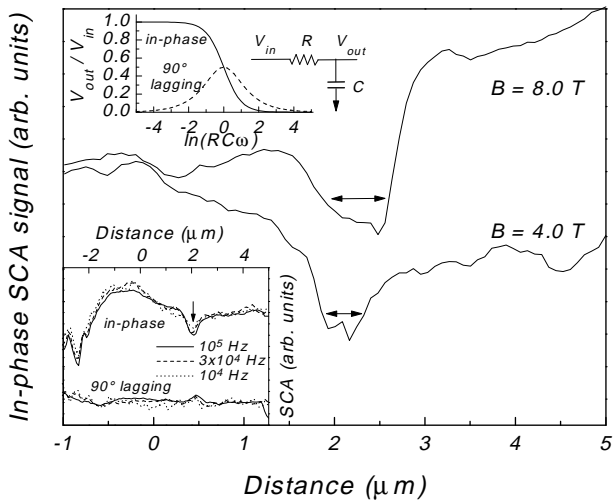}
\caption{\narrowtext
The SCA signal across the $\nu = 4$ and $\nu = 2$
strips. The curves are offset for clarity. The data are taken at the
same conditions as in Fig. 1. Lower inset: the in-phase and $90
^\circ$ lagging SCA signals at different frequencies around $\nu = 4$.
The vertical scale is the same for all frequencies. The $90 ^\circ$
lagging signals are arbitrarily offset. Upper inset: a model diagram
of the 2DEG charging. }
\end{figure}

Theory predicts that magnetic field should modify the electron
distribution, so that within the incompressible strip the electron
density is everywhere fixed at a level corresponding to an integer
Landau level filling factor\cite{Chklovskii}. This picture relies on
having zero DOS in the cyclotron gap and it requires
alteration if the DOS is nonzero. The resulting
screening should eliminate the region of $\frac{dn}{dx}=0$.
Nevertheless, even in this case a low compressibility strip should be
formed in the regions where the Fermi energy lies between the Landau
levels and the DOS is low. If the average DOS between the Landau
levels is $D$, then the areal density of
these states is $\sim \hbar \omega_{c} D$. This quantity should be
equal to the span of the electron density across the strip
$\frac{dn}{dx}w$. We estimate the average DOS between
the Landau levels as $D \sim \frac{dn}{dx}w / \hbar \omega_{c}$.
Taking the strip width $w \sim 0.5 \mu m$, we obtain an approximately
tenfold suppression of the DOS as compared to the zero field value.
This result concurs with our estimate based on the signal strength
above. The corresponding screening radius, $\frac{\epsilon}{2 \pi
e^{2} D}\sim 50nm$ is much smaller than the width of the strip. Thus,
the basic assumption of Ref.\cite{Chklovskii} that the screening by
the states in the cyclotron gap is negligible, does not hold in our
case. 

Larkin and Levitov considered the broadening of the low
compressibility strip as a result of a nonzero DOS
between Landau levels\cite{Larkin}. They found that the width of the
strip is quite accurately given by $w=w_{0}+\Delta n/\frac{dn}{dx}$,
where $\Delta n$ is the areal density of electrons in the cyclotron
gap. This expression shows that when $w \gg w_{0}$, we may approximate
$w$ by $\Delta n/\frac{dn}{dx}$, as performed above. 

Efros has investigated theoretically a quantum Hall liquid in the
presence of a macroscopic density gradient and random potential due to
the ionized donors\cite{Efros1}. He predicts an areal density of
electron states in the cyclotron gap of $2 n_{c}$, where $n_{c} =0.4
\sqrt{C}/s$, $C$ is the density of the ionized donors and $s=20nm$ is
the spacer between them and the 2DEG. Correlations among the ionized
donors are known to reduce the amplitude of the random potential, so
that $C$ should be substituted by an effective $C_{0} \sim 4 \times
10^{10}/cm^{2}$\cite{Efros2}. This gives $2 n_{c} \approx 8 \times
10^{10}/cm^{2}$. We estimate from the experiment a close value of $w
\frac{dn}{dx} \lesssim 5 \times 10^{10}/cm^{2}$. This areal density of
states in the gap between Landau levels also agrees well with the one
measured in a sample of a similar quality at $\nu=4$ in Fig. 3 of
Ref.\cite{Wei}.

In summary, we have studied the 2DEG in the Quantum Hall regime close
to integer filling factors. In a smooth potential gradient, we observe
low compressibility strips corresponding to integer Landau level
filling factors. Our measurements indicate that the strips are
significantly wider than predicted by theory not including a
short-range disorder potential\cite{Chklovskii}. We conclude that in
our sample the width of the low compressibility strips at integer
filling factors is determined by a low, but nonzero density of states
between the Landau levels.

We thank L.S. Levitov for crucial discussions. This work was supported
by the Office of Naval Research, the Packard Foundation, JSEP, and the
National Science Foundation DMR.

\figure Fig. 1: The SCA measurements at different magnetic fields
around $\nu =4$ across the area of enhanced density. The curves are
offset for clarity. Left inset: the Kelvin probe signal. Right inset:
a schematic description of the SCA technique. 

\figure Fig. 2: $13 \times 13 \mu m$ SCA images at different magnetic
fields near $\nu=2$ presented on a gray scale. The left and right
images correspond to the in-phase and $90 ^\circ$ lagging SCA signals,
respectively. The measurements are performed at magnetic fields of a)
$B=8.0$ T, b) $B=8.1$ T and c) $B=8.2$ T. Fig. 2a and c represent the
limiting cases where the SCA signal is dominated by compressibility or
resistivity, respectively. The data are taken at a different location
than Fig. 1.

\figure Fig. 3: The SCA signal across the $\nu = 4$ and $\nu = 2$
strips. The curves are offset for clarity. The data are taken at the
same conditions as in Fig. 1. Lower inset: the in-phase and $90
^\circ$ lagging SCA signals at different frequencies around $\nu = 4$.
The vertical scale is the same for all frequencies. The $90 ^\circ$
lagging signals are arbitrarily offset. Upper inset: a model diagram
of the 2DEG charging. 

\end{multicols} 


\begin{references}
\bibitem{comment} Some signal is detected even when the scanning probe
is located above a non charging region, due to the stray capacitance
of the contacting wires and the rest of the 2DEG to the scanning
probe. We carefully balance this signal away by a standard capacitor.
\bibitem{comment2} The use of the parallel plate capacitor model is
partly justified by an inverse proportionality of the SCA signal on
the tip-sample separation $h$, observed for $h \lesssim 100
nm$. However, the value of $h$ is not known precisely,
and this makes the estimate of $D$ accurate only up to an order of
magnitude. Alternatively one might model the tip by a thin rod
or a point-like object. It may be shown that in these cases the
contrast of the SCA signal above the low compressibility strip is
determined primarily by the spatial motion of the strip in response to
the AC excitation. This would result in a SCA contrast much smaller
than observed in the experiment, which further justifies our use of
the parallel plate approximation for the scanning tip.
\bibitem{comment3} Alternatively, an apparent broadening of the strip
might be caused by poorly charging localized states on the edges of
the strip, as suggested in Ref.\cite{Chklovskii}. However, this effect
would result in a phase lagging signal, which is absent in Figure 3.
\end{references}
\end{document}